\documentstyle[12pt]{article}
\begin{document}

\title{Scaling Behavior of Granular Particles \\
in a Vibrating Box}
\author{Jysoo Lee\\
Benjamin Levich Institute and Department of Physics \\
City College of the City University of New York \\
New York, NY 10031}
\date{\today}
\maketitle

\begin{abstract}

Using numerical and analytic methods, we study the behavior of
granular particles contained in a vibrating box. We measure, by
molecular dynamics (MD) simulation, several quantities which
characterize the system. These quantities---the density and the
granular temperature fields, and the vertical expansion---obey scaling
in the variable $x = Af$. Here, $A$ and $f$ are the amplitude and the
frequency of the vibration. The behavior of these quantities is
qualitatively different for small and large values of $x$. We also
study the system using Navier-Stokes type equations developed by Haff.
We develop a boundary condition for moving boundaries, and solve for
the density and the temperature fields of the steady state in the
quasi-incompressible limit, where the average separation between the
particles is much smaller than the average diameter of the particles.
The fields obtained from Haff's equations show the same scaling as
those from the simulations. The origin of the scaling can be easily
understood. The behavior of the fields from the theory is consistent
with the simulation data for small $x$, but they deviate significantly
for large $x$. We argue that the deviation is due to the breakdown of
the quasi-incompressibility condition for large $x$.

\vspace{14pt}
\end{abstract}

\newpage
\section{Introduction}
\label{sec:intro}

Systems of granular particles (e.g. sand) exhibit many interesting
phenomena, such as segregation under vibration or shear, density waves
in the outflow through a hopper and a tube, and the formation of heaps
and convection cells under vibration \cite{s84,c90,jn92,m92}. These
phenomena are consequences of the unusual dynamical response of the
systems, many of which are still poorly understood.

Granular particles in motion lose energy due to inelastic collisions
of the particles. Thus energy has to be supplied to granular systems
from an external source(s) in order to sustain the movement of the
particles. We can classify granular systems according to the way in
which energy is supplied, or how the systems are agitated. A few
common ways to agitate the systems are shear, vibration and body force
(e.g.  gravity). Here, we consider only vibrational agitation, which
is probably one of the most popular method. There are many interesting
phenomena associated with the vibrated systems, such as convection
cells \cite{zs91,t92a,ghs92,lcbrd94a}, heap formation
\cite{f31,er89,ldf89,cdr92,l94}, subharmonic instability \cite{dfl89},
surface waves \cite{pb93,mus94} and even turbulent flows \cite{t92b}.
The basis for understanding these diverse phenomena is, in our
opinion, to understand of the state of granular media under vibration.
The state is characterized by several fields of the system, for
example, the density, velocity and granular temperature fields.

There have been several studies on the state of granular systems under
vibration. Thomas {\em et al} studied the system in three dimensions,
mainly focusing on the behavior of shallow beds \cite{tmls89}.
Cl\'{e}ment and Rajchenbach experimentally measured the density,
velocity and temperature fields of a two-dimensional vertical packing
of beads \cite{cr91}. They found that the temperature increases
monotonically as the distance from the bottom plate is increased. The
same system was studied by molecular dynamics (MD) simulation with
similar results \cite{ghs92b}. In a series of simulations and
experiments, Luding {\em et al} studied the behavior of the one and
two dimensional systems \cite{lcbrd94b,lhb94,lt}. They found that the
vertical expansion, which is the increase of the center of mass due to
the vibration, scales in the variable $x = Af$. Here, $A$ and $f$ are
the amplitude and the frequency of the vibration. The expansion
behaves as $x^{2}$ and $x^{3/2}$ in one and two dimensions,
respectively. The reported shape of the temperature fields is {\em
decreasing} as the distance from the bottom is increased, in contrast
to the previous result \cite{lcbrd94b,lt}. In a recent MD simulation
of the three dimensional system, Lan and Rosato measured the density
and temperature fields \cite{lr94}.  They compared the results with
the theoretical predictions by Richman and Martin \cite{rm92}, and
found good agreements. More discussion on the theory will come later.
They also found the shape of the temperature field is different for
low and high amplitude of the vibration. Finally, an approximate
theory was developed for the system in one dimension, which agrees
with simulations in the weak and the strong dissipative regime
\cite{mbd94}.

In this paper, we measure the density, the granular temperature and
the expansion for the two dimensional system using MD simulation. We
find that not only the expansion, but also the density and the
temperature fields, scale in the variable $x$. The behavior of these
quantities is qualitatively different for small and large values of
$x$. The different shapes of temperature field reported can be due to
different values of $x$ used in the measurements. We also study the
system using continuum equations developed by Haff \cite{h83}. We
develop a boundary condition for moving boundaries. We solve the
equations for the density and the temperature fields and the expansion
of the steady state. We consider only the quasi-incompressible case,
where the average separation between the particles is much smaller
than the average diameter of the particles. The fields obtained from
Haff's equations show the same scaling as those from the simulations.
Furthermore, the origin of the scaling can be traced back to the
boundary conditions, and can be easily understood.  Also, behaviors of
the fields from the theory are consistent with the simulational data
for small $x$, but they deviate significantly for large $x$. We argue
that the discrepancy is due to the breakdown of the
quasi-incompressibility condition for large $x$. Some of these results
can also be obtained using the theory by Richman and Martin, which is
based on an extension of the Boltzman equation for inelastic gas
\cite{sj81,js83}.

The paper is organized as follows. In Sec.~\ref{sec:sim}, we start
from defining the interaction of the particles used in the MD
simulations. The geometry and various fields of the system will be
discussed. We then present the expansion and the fields obtained by
the simulations. Analytic results will be discussed in
Sec.~\ref{sec:ar}. The continuum equations for granular material will
be given, and simplified for the present system. We then derive a
boundary condition for moving walls, and present the solution of the
equations. In Sec.~\ref{sec:comp}, we compare the fields obtained by
the simulations with those by the theory. Finally, conclusions are
given in Sec.~\ref{sec:con}.

\section{Molecular Dynamics Simulations}
\label{sec:sim}

\subsection{Interaction of the Particles}
\label{sec:pint}

We start by describing the interactions used in the molecular dynamics
simulations. The force between two particles $i$ and $j$, in contact
with each other, is the following. Let the coordinates of the center
of particle $i$ ($j$) be $\vec{R}_i$ ($\vec{R}_j$), and $\vec{r}
\equiv \vec{R}_i - \vec{R}_j$. In two dimensions, we use a new
coordinate system defined by two vectors $\hat{n}$ (normal) and
$\hat{s}$ (shear). Here, $\hat{n} = \vec{r} / {\vert \vec{r} \vert}$,
and $\hat{s}$ is defined as rotating $\hat{n}$ clockwise by $\pi/2$.
The normal component $F_{j \to i}^{n}$ of the force acting on particle
$i$ by $j$ is
\begin{equation}
\label{eq:fn}
F_{j \to i}^{n} = k_n (a_i + a_j - \vert \vec{r} \vert)
                - \gamma_n m_e (\vec{v} \cdot \vec{n}),
\end{equation}
where $a_i$ ($a_j$) is the radius of particle $i$ ($j$), and $\vec{v}
= d\vec{r}/dt.$ The first term is an linear elastic force, where $k_n$
is the elastic constant of the material. The constant $\gamma_n$ of
the second term is the friction coefficient of a velocity dependent
damping term, $m_e$ is the effective mass, $m_i m_j/(m_i + m_j).$ The
shear component $F_{j \to i}^{s}$ is given by
\begin{equation}
F_{j \to i}^{s} = - \gamma_s m_e (\vec {v} \cdot \vec {s}) - {\rm
sign} (\delta s) ~ {\rm min}(k_s \vert \delta s \vert, \mu \vert F_{j
\to i}^n \vert),
\label{eq:fs}
\end{equation}
where the first term is a velocity dependent damping term similar to
that of Eq.~(\ref{eq:fn}). The second term is to simulate static
friction, which requires a {\em finite} amount of force ($\mu F_{j \to
i}^{n}$) to break a contact \cite{cs79}. Here, $\mu$ is the friction
coefficient, $\delta s$ the {\em total} shear displacement during a
contact, and $k_s$ the elastic constant of a virtual spring. There are
several studies on granular systems using the above type of
interactions \cite{gramd}. However, only a few of them
$\cite{cs79,gramd,bg91}$ include static friction. A particle can also
interact with a wall. The force on particle $i$, in contact with a
wall, is given by Eqs.~(\ref{eq:fn})-(\ref{eq:fs}) with $a_j =
\infty$ and $m_e = m_i$.  A wall is assumed to be rigid, i.e. it is
not moved by collisions with the particles. Also, the system is under
a gravitational field $\vec{g}$. We do not include the rotation of the
particles in present simulation. A detailed explanation of the
interaction is given elsewhere \cite{lh93}.

The trajectories of the particles are calculated using a fifth order
predictor-corrector method. The interaction parameters used in this
study are fixed as follows, unless otherwise specified. They are $k_n
= 5 \times 10^6, k_s = 1 \times 10^4, \gamma _n = 1 \times 10^3,
\gamma _s = 0$ and $\mu = 0.2$. The timestep is taken to be $5
\times 10^{-6}$. This small timestep is necessary for the large elastic
constant used in the simulations. For too small values of the elastic
constant, the system loses the character of a system of distinct
particles, and behaves like a viscous material. In order to avoid
artifacts of monodisperse systems (e.g., hexagonal packing), we choose
the radius of the particles from a Gaussian distribution with the mean
$0.1$ and the width $0.02$. The density of the particles is $0.5$.
Throughout this paper, CGS units are implied.

\subsection{Setup for the Simulations}
\label{sec:ssetup}

We put the particles in a two-dimensional rectangular box. The box
consists of two horizontal (top and bottom) plates which oscillate
sinusoidally along the vertical direction with a given frequency and
amplitude. The separation between the two plates is chosen to be much
larger ($10^5$ times) than the average radius of the particles, so the
particles never interact with the top plate for all the cases studied
here. For simplicity, we apply a periodic boundary condition in the
horizontal direction.

We start simulation by inserting the particles at random positions in
the box. We let them fall by gravity and wait until they lose energy
by collisions. We wait for $10^5$ iterations for the particles to
relax, and during this period we keep the plates fixed.  The typical
velocity at the end of the relaxation is of order of $10^{-2}$. After
the relaxation, we start oscillating the plates. We vibrate for about
$50$ cycles before taking any measurements in order to eliminate any
transient effect.

The main quantities we are interested in are density, velocity and
granular temperature, which probably are the most basic quantities to
characterize the system. In order to measure these fields, we divide
the system into rectangular cells of width $w$ and height $h$. For
cell $i$, we identify the particles whose center lies within the cell,
which we label by $j = 1, 2, 3 ..., N_i$. Here, $N_i$ is the total
number of particles in the cell. The density at cell $i$, $\rho_i$, is
defined as
\begin{equation}
	\rho_i \equiv {1  \over w \cdot h} ~\sum_{j=1}^{N_i} \pi r_j^2,
	\label{defrho}
\end{equation}
where $r_j$ is the radius of particle $j$. We define an average of a
certain quantity $A$ in cell $i, \langle A \rangle_i$, as
\begin{equation}
	\langle A \rangle_i \equiv {\sum_{j=1}^{N_i} m_j A_j \over
	                    \sum_{j=1}^{N_i} m_j},
	\label{defavg}
\end{equation}
where $m_j$ is the mass of particle $j$. Then, the velocity field at
cell $i, \vec{v_i}$, is given by
\begin{equation}
    \vec{v_i} \equiv  \langle \vec{v} \rangle_i,
	\label{defvel}
\end{equation}
and the granular temperature at cell $i, T_i$ is
\begin{equation}
    T_i \equiv {1 \over 2}~
               (\langle v_x^2 - \langle v_x \rangle_i^2 \rangle_i
        +       \langle v_y^2 - \langle v_y \rangle_i^2 \rangle_i).
	\label{defgrt}
\end{equation}
We also measure the vertical component of the center of mass $y_{cm}$
of the particles, which is an useful quantity to compare with
theoretical predictions. In principle, we can calculate $y_{cm}$ from
the measured density fields, but the calculation introduces additional
sources of uncertainty. We measure the above quantities by averaging
over at least three independent runs, where each run is averaged over
about $200$ cycles of vibrations.

We study the effects of the width $W$ of the vibrating plates on the
measured quantities in order to determine the optimal width for the
main measurements. We measure the density and the granular temperature
fields for $W = 1, 2, 3, 4$, where the height $H$ of the pile is
roughly fixed constant by changing the number of particles to be $50,
100, 150, 200$. The resulting height is little smaller than $2$. Also,
we use $w = W$ and $h = 0.2$. Here, the large $w$ is chosen due to the
translational symmetry resulted from the periodic boundary condition.
We do not find any systematic dependence of the fields on $W$ except
that they become less noisy for larger $W$, which is simply due to the
larger number of particles contained in a cell. We find $W = 1$ is a
good compromise between the quality of data and the computation time.
The dependence of the fields on $H$ will be discussed later.

\subsection{Measurement of the Expansion}
\label{sec:mexp}

We now discuss the quantities obtained as described previously. We
start from the vertical expansion of the pile $y_{exp}$, defined as
the difference in the vertical center of mass $y_{cm}$ during and
before vibration. Since the expansion can be expressed as an integral
involving the density field, $y_{exp}$ can be used as a number
representative of the density field. In Fig.~1, we show $y_{exp}$
measured for various values of the amplitudes $A$ and the frequencies
$f$ of the vibration \cite{timestep}. Here, the width $W = 1$ and the
number of particles $N = 50$. Note that the expansion is plotted
against a scaling variable $x = A f$. For the entire range of variable
$x$, we do not find any systematic deviation from the scaling behavior
except that the lowest frequency data ($f = 20$) seems to be slightly
off. This scaling behavior was first measured in the simulations by
Luding {\em et al} \cite{lcbrd94b,lhb94,lt}. For intermediate values
of $x$, our data is also consistent with the $x^{3/2}$ behavior
proposed by Luding {\em et al} \cite{lhb94}. However, for small and
large values of $x$, our data shows deviations from the behavior. The
pattern of the deviation is not affected by changing the values of the
elastic constant $k_n$ (to $10^6$ and $5 \times 10^5$). The scaling
behavior and the deviation will later be discussed in more detail.

Since we are concerned about the possibility that $y_{cm}$ is
dominated by a few top particles isolated from the main part of the
pile, we also try an alternative definition of the expansion. We sort
the particles in a box according to their vertical coordinates. We
then define $y_{center}$ as the vertical coordinate of the center of
the particle which is the $(N/2)$-th in the list. Here, $N$ is the
total number of particles in the box. The quantity $y_{center}$ still
contains the meaning of the center of mass, but is now not dominated
by a few particles. We find that, in contrast to our worries, the
expansions using $y_{center}$ behave essentially the same way as those
using $y_{cm}$.  We thus use the definition using $y_{cm}$ from now
on.

\subsection{Measurement of Density and Temperature}
\label{sec:mfield}

We discuss the density $\rho(y)$ and the granular temperature $T(y)$
fields. First, we study the behavior of the fields at the bottom
plate, i.e., in the lowest cell ($y=0)$. These values are, as we shall
see later, easier to compare with theoretical predictions than the
entire fields. In Fig.~2 (a), we show the density $\rho (y=0)$ vs $x$
for various values of $A$ and $f$. Here, we find the same scaling
behavior as the expansion $y_{exp}$ (Fig.~1). This is not surprising
considering the relation between the expansion and the density field.
We also find the same scaling behavior for the temperature. The square
root of $T(y=0)$ is plotted against the scaling variable $x$ in
Fig.~2(b). Although there are larger fluctuations, the same scaling is
still apparent. Also, the decay of $T(y=0)$ for large $x$ is partly
due to a dip in the density near the bottom plate.

We then proceed to study the entire $y$-dependent fields. In Fig.~3,
we show the density field $\rho (y)$ for several values of $A$ and
$f$. Here, the value of $x$ is fixed to be $3$ and $10$ in Fig.~3(a)
and (b), respectively. In the figures, the quality of scaling is
poorer than those shown previously (Figs.~1 and 2). However, the
quality of scaling will be improved by discarding the lowest frequency
data $(f = 20)$. The density field plotted against $y / y_{exp}$ was
also shown to collapse into a single curve \cite{lt}, which is
consistent with above scaling, since $y_{exp}$ scales in $x$. The
situation is not much different for the temperature field $T(y)$. In
Fig.~4(a) and (b), we show the square root of $T(y)$, where $x$ is
again fixed to be $3$ and $10$, respectively. Here also, the quality
of scaling is not great, and will be improved by ignoring the $f = 20$
data. Putting aside the deviations at low frequencies, it is not
entirely clear from these data that either the density or the
temperature field exhibits more than an approximate scaling.

One interesting point is the shape of the temperature field. For large
values of $x$, the field $T(y)$ is a monotonically decaying function
of the height $y$ (as shown in Fig.~4(b)). As one decreases $x$, a
local maximum begins to appear. In Fig.~4(a) one can see an maximum is
starting to form around $y = 2$. As one decreases $x$ further, the
maximum becomes more and more pronounced. The above observation can
explain the different shapes of $T(y)$ reported by several experiments
and simulations on the granular systems. In experiments of the two
dimensional system, a local maximum is always present in the $T(y)$
field \cite{cr91}. In contrast, $T(y)$ is found to be monotonically
decaying in simulations of the one dimensional systems
\cite{lcbrd94b,lt}. Other simulations of the system in three
dimensions also report monotonic decay of the temperature field. In
the three dimensional simulations, however, changes of the shape of
the field for smaller amplitudes were noted, although were not
systematically studied \cite{lr94}.

We conclude this section by presenting the dependence of $y_{exp}$ on
the total height $H$ of the pile. We change the total height by
varying the number of particles $N$ and keeping the width $W$ fixed.
We keep the previous value of $W = 1$. In Fig.~5, we show the scaled
expansion for $N = 50, 100, 150$ and for several values of $A$, where
$f$ is fixed to $100$. The expansion is scaled by $50 / N$, where the
factor $50$ is used to keep the $N = 50$ data unchanged. In the
figure, we find a reasonable collapse of the curves. There seems to be
no systematic deviation, although the data for large $x$ seems to be
slightly off. The scaling is first found by Luding {\em et al}
\cite{lcbrd94b}.

In this section, we have presented the data from the MD simulations.
The expansion as well as the density and the granular temperature
fields exhibit scaling in the variable $x$, but the quality of the
scaling for the fields is worse than that of the expansion. Also, the
lowest frequency data seems to deviate from the scaling. In the next
section, we present a continuum theory of granular material which can
explain the origin of the observed scaling.

\section{Analytic Results}
\label{sec:ar}

\subsection{Basic Equations}
\label{sec:be}

The system of granular particles needs an external agitation(s) in
order to sustain the movement of the particles. Without such an
agitation, the motion of the particles will be suppressed because of
the constant energy loss resulting from inelastic collisions of the
particles. Many granular systems with an external agitation, including
the system of granular particles in a vibrating box, can loosely be
described as inelastic gas. A particle in such a system constantly
moves. It assumes a ballistic trajectory until it collides with
another particle or, if there is, a wall of the container. After the
collision, the particle again moves ballistically. Such movement is
analogous to those of particles in a gas. Motivated by the analogy,
tools used to develop the kinetic theory of gas have been adapted to
study granular systems. One such effort is to extend the Boltzman
equation to an inelastic gas \cite{sj81,js83}. In this approach, the
time evolution of the fields (velocity, density and granular
temperature) is written in terms of an integral involving all possible
collisions, namely the Boltzman integral. Another approach, due to
Haff, is to develop equations of motion very similar to the
Navier-Stokes equations \cite{h83}. However, the coefficients in the
equations, e.g. viscosity, are no longer constant, but now functions
of the fields. Both approaches are not yet complete, and they contain
assumptions whose validities have yet to be checked. Haff's approach
seems to be less rigorous; for example, the dependence of the
coefficients on the fields are derived based only on intuitive
arguments. On the other hand, the equations one has to solve in Haff's
approach are much simpler. Despite its simplicity, known solutions for
a few systems using Haff's approach do not significantly differ from
those using the other method \cite{h83,js83}. In this paper, we will
follow Haff's approach.

The equations of motion developed by Haff consist of mass, momentum
and energy conservation. The mass conservation equation is
\begin{equation}
\label{eq:mass}
{\partial \over \partial t} \rho + \vec{\nabla} \cdot (\rho \vec{v})
= 0,
\end{equation}
where $\rho$ and $\vec{v}$ are the density and the velocity fields,
respectively. The equation (\ref{eq:mass}) is exactly the same as that
of the Navier-Stokes equations. Next is the $i$-th component of the
momentum conservation equation,
\begin{equation}
\label{eq:mom}
\rho {\partial \over \partial t} v_i + \rho (\vec{v} \cdot
\vec{\nabla}) v_i = {\partial \over \partial x_i} [ -p + \lambda
(\vec{\nabla} \cdot \vec{v})] + {\partial \over \partial x_j} [\eta
({\partial v_j \over \partial x_i} + {\partial v_i \over \partial
x_j})] + \rho g_i,
\end{equation}
where summation over index $j$ is implied. The coefficients $\lambda$
and $\eta$ are viscosities which will be determined later. Also, $p$
is the internal pressure, and $g_i$ is the $i$-th component of the
gravitational field. The form of Eq. (\ref{eq:mom}) is again the same
as that of the Navier-Stokes equations. However, the coefficients as
well as the internal pressure are now functions of the fields instead
of being constant. The last of the equations of motion is energy
conservation,
\begin{eqnarray}
\label{eq:energy}
{\partial \over \partial t} ({1 \over 2} \rho v^2 + {1 \over 2}
\rho T) & + & {\partial \over \partial x_i}
[( {1 \over 2} \rho v^2 + {1 \over 2} \rho T) v_i] \\ \nonumber
& = & -{\partial \over \partial x_i} (p v_i) \\ \nonumber
& + & {\partial \over \partial x_i}[\lambda (\vec{\nabla} \cdot
\vec{v}) v_i] + {\partial \over \partial x_i}[\eta ({\partial v_i
\over \partial x_j} + {\partial v_j \over \partial x_i}) v_j] \\
\nonumber & + & \rho v_i g_i \\ \nonumber
& + & {\partial \over \partial x_i} [K {\partial \over
\partial x_i} ({1 \over 2} \rho T)] - I.
\end{eqnarray}
Here, $T$ is the granular temperature field defined in Eq.
(\ref{defgrt}), $K$ is the ``thermal conductivity,'' and $I$ is the
rate of dissipation due to inelastic collisions \cite{t}. Also, the
summation over repeated indices is implied. Although the form of Eq.
(\ref{eq:energy}) is somewhat different from that of the Navier-Stokes
equations, the equation can still be easily understood. The left hand
side of Eq. (\ref{eq:energy}) is simply the material derivative of the
total kinetic energy, where the total kinetic energy is divided into
the convective part (involving $\vec{v}$) and the fluctuating part
(involving $T$). On the right hand side of the equation, the first
three lines are simply the rate of work done by the internal pressure,
viscosity and gravity, respectively. The term involving $K$ is the
rate of energy transported by ``thermal conduction.'' The term $I$,
which is a consequence of inelasticity of the particles, is
responsible for many unique properties of granular material.

We now discuss the above coefficients which are yet to be determined.
The relations of these coefficients to the fields are derived based on
intuitive arguments. Also, the derivation assumes that the density is
not significantly smaller than the closed packed density, i.e., the
system is almost incompressible. The relation for the internal
pressure is
\begin{equation}
\label{eq:intp}
p = t d \rho {T \over s},
\end{equation}
where $t$ is an undetermined constant, $d$ is the average diameter of
the particles. The variable $s$, which is roughly the gap between the
particles, is related to the density by
\begin{equation}
\label{eq:sandr}
\rho \equiv {m \over (d + s)^3},
\end{equation}
where $m$ is the average mass of the particles. Then, the viscosity
$\eta$ is given as
\begin{equation}
\label{eq:eta}
\eta = q d^2 \rho {\sqrt{T} \over s},
\end{equation}
where $q$ is an undetermined constant. In a similar way, the thermal
conductivity is found to be
\begin{equation}
\label{eq:K}
K = r d^2 {\sqrt{T} \over s}.
\end{equation}
Here again, $r$ is an undetermined constant. Finally, the rate of
dissipation is
\begin{equation}
\label{eq:I}
I = \gamma \rho {T^{3/2} \over s}.
\end{equation}
Also, $\gamma$ is an undetermined constant. One can notice the
viscosity $\lambda$ is left undetermined. This is due to the fact that
in the range where these relations are valid, the term containing
$\lambda$ is negligible, and is dropped from the calculation.

\subsection{Equations for Particles in a Vibrating Box}
\label{sec:evb}

We now apply Haff's equations to the system being considered in the
present paper---granular particles under vibration. Since the general
solution of the problem using Haff's equations are too difficult to
obtain analytically, we introduce several constraints which simplify
the equations. First, we only study steady state properties of the
system. In order words, we are interested in only the time averaged
quantities. Since there is no net flow of particles in the steady
state, the time averaged values of $\vec{v}$ is zero, which greatly
simplifies the equations. The second simplification is resulting from
the horizontal periodic boundary conditions we imposed.  Due to the
boundary condition, there is no significant variation of the fields
along the horizontal direction. As a result we only have to deal with
a one dimensional equation instead of two or three dimensional ones.
The third condition is incompressibility, which is a little bit tricky.
Incompressibility implies, strictly speaking, that the density $\rho$
is constant. Due to the relation between $\rho$ and $s$ (Eq.
(\ref{eq:sandr})), $s$ also has to be constant.  Here, we are
interested in the situation where $s$ is much smaller than $d$, but
still non-zero. In such cases, the variation of the density can be
ignored, but not the variations of the variables that depends directly
on $s$. We call this condition quasi-incompressibility.

The simplified equations for the pile of granular particles under
vibration are then
\begin{equation}
\label{eq:nmass}
{d \over d y} v = 0,
\end{equation}
\begin{equation}
\label{eq:nmom}
{d \over dy} p + \rho g = 0,
\end{equation}
and
\begin{equation}
\label{eq:nenergy}
{1 \over 2} \rho {d \over dy}[K {d \over dy} T]  - I = 0.
\end{equation}
Here, all the fields are averaged over time, so they are not functions
of time, but of spatial coordinate $y$ only. We thus replace partial
derivatives in the equations with ordinary derivatives. Note that the
first equation is automatically satisfied. Substituting the relation
for the internal pressure Eq.~(\ref{eq:intp}) to Eq.~(\ref{eq:nmom}),
we obtain
\begin{equation}
{d \over dy} ({T \over s}) = -{g \over td},
\end{equation}
whose solution is
\begin{equation}
\label{eq:solts}
{T \over s} = -{g \over td} (y - y_o),
\end{equation}
where $y_o$ is a constant which will be discussed later. Substituting
Eq. ~(\ref{eq:solts}) to Eq.~(\ref{eq:nenergy}), we obtain
\begin{equation}
\label{eq:st1}
{d^2 \over dy^2} \sqrt{T} + {1 \over y - y_o} ~ {d \over dy}
\sqrt{T} - {\gamma \over rd^2} \sqrt{T} = 0,
\end{equation}
which is a linear differential equation for $\sqrt{T}$. We make a
change of variable,
\begin{equation}
z \equiv {y- y_o \over \ell},
\end{equation}
where $\ell$ is defined as $\sqrt{r / \gamma} ~ d$. By inserting the
new variable to Eq.~(\ref{eq:st1}), we arrive at the final form of the
equation
\begin{equation}
\label{eq:final}
{d^2 \over dz^2} \sqrt{T} + {1 \over z} {d \over dz} \sqrt{T}
- \sqrt{T} = 0.
\end{equation}
The solutions the equation are the modified Bessel functions $I_o(z)$
and $K_o(z)$. However, $K_o(z)$ has a singularity at the physical
region, and can not be a part of the solution. The solution is,
therefore,
\begin{equation}
\label{eq:sol}
\sqrt{T} = A ~ I_o(z),
\end{equation}
where $A$ is a constant yet to be determined. The result obtained so
far is identical to that of Haff \cite{h83}. We now discuss our
contribution to the problem, which concerns the boundary conditions.

\subsection{Boundary Condition}
\label{sec:bc}

One of the areas which are yet to be completed in Haff's theory is
boundary conditions. For example, what is the boundary condition one
has to impose for the temperature field $T$ and the velocity field
$\vec{v}$ at the bottom of a vibrating box? The boundary conditions
for fixed walls have been developed based on energy conservation
across the boundary \cite{hhuj84}. We will show that a boundary
condition for moving walls can also be derived from the similar
condition of energy conservation. As noted in \cite{hhuj84}, there are
two energy flows near a boundary. One is the thermal conduction, and
the other is the energy transfer by collisions between the particles
and the boundary.  Energy conservation implies that the rate of energy
transferred by the conduction has to be equal to that by the
collisions. We consider an horizontal boundary vibrating vertically,
like the bottom plate of a vibrating box. The rate of the thermal
conduction per area at the boundary is given as
\begin{equation}
\label{eq:thrm}
{1 \over 2} K \rho {d \over dy} T \mid _{y = 0},
\end{equation}
where $y=0$ is the average vertical coordinate of the boundary. To
calculate the rate of energy transferred by the collisions, consider a
particle colliding with the boundary. If convection is absent, the
movement of the particle is resulted from the fluctuation. The typical
velocity due to the fluctuation is $\sqrt{T}$. The velocity of the
particle after the collision will depend on the velocity of the
boundary at the moment of the collision.  Here, we assume that the
phase of the vibration at which the collision occurs is random---an
assumption whose validity will be discussed later. In such cases, the
typical velocity of the wall is $v_w = A f$.  The amount of energy
transferred in one collision is
\begin{equation}
\label{eq:1col}
{1 \over 2} m[ T (1 - e_w^2) - (1 + e_w)^2 v_w^2],
\end{equation}
where $e_w$ is the coefficient of restitution between a particle and
the wall. The rate of energy transfer per area is Eq.~(\ref{eq:1col})
multiplied by the frequency of collisions $T / s$ divided by the area
of contact $d^2$. Here, we again use the quasi-incompressibility.
Equating this rate to Eq.~(\ref{eq:thrm}), and use the definition of
$K$ (Eq.~(\ref{eq:K})), we obtain
\begin{equation}
\label{eq:bc}
T (1 - e_w^2) - (1 + e_w)^2 v_w^2 = {rd \over a} {d \over dy} T,
\end{equation}
where $a$ is an undetermined constant. The method used to derive the
above boundary condition can be extended to a case where the boundary
is moving horizontally, and even to a case with both horizontal and
vertical motions.

We now apply the above boundary condition to the problem of the
vibrating box. First, note that the modified Bessel function $I_o(z)$
can be approximated as
\begin{equation}
\label{eq:asymi}
I_o(z) \sim {\exp (z) \over \sqrt{2 \pi z}},
\end{equation}
for $z > 0$. The expression is actually an asymptotic form of $I_o(z)$
for $z \gg 1$, but it still is a good approximation even for small
values of $z$. By applying the boundary condition Eq.~(\ref{eq:bc}) to
the solution Eq.~(\ref{eq:sol}) with the approximation
Eq.~(\ref{eq:asymi}), we obtain
\begin{eqnarray}
A^2 & = & {1 \over I_o^2(-y_o/\ell)} ~ {(1 + e_w)^2
\over 1 - e_w^2 - (2rd / a \ell) ~ (1 - \ell / 2 y_o)} ~
v_w^2 \\ \nonumber
& \equiv & {1 \over I_o^2(-y_o/\ell)} ~ B^2 ~ v_w^2,
\end{eqnarray}
where a new variable $B$ is introduced. Substituting for $A$ in the
solution Eq.~(\ref{eq:sol}), we now arrive at the final form of the
solution for the temperature field
\begin{equation}
\label{eq:tsol}
\sqrt{T(y)} \simeq B ~ v_w ~ \sqrt{y \over y_o - y} ~ \exp (-y/\ell),
\end{equation}
where we use the property $I_o(-z) = I_o(z)$. Besides the prefactor
$B$, the above result is the same as that obtained by Haff who simply
assumed that $\sqrt{T}$ at the boundary is $v_w$ \cite{h83}.  The
temperature field decays exponentially with decay length $\ell$ as one
moves away from the bottom plate. Note that it has a singularity at $y
= y_o$, which is an artifact resulting from the fact that Haff's
theory is not valid near the top part of the pile. In other words, the
theory is based on the picture that the state of the system is
gaseous.  Near the top of the pile, however, the particles do not
collide frequently, but assume ballistic trajectories \cite{h83}. On
the other hand, Eq.~(\ref{eq:sol}) is still valid for $y$ not very
close to $y_o$, since the gaseous picture is valid for that ranges of
$y$. Thus, the behavior of $T$ to increase again for large $y$ is a
physical effect, not an artifact. A word about $y_o$.  One can see
$y_o$ is still left undetermined. According to Eq.~(\ref{eq:solts}),
the condition to determine $y_o$ is that the internal pressure
vanishes at $y= y_o$.  However, since the solution we have obtained is
not valid at $y = y_o$, we can not impose the condition to determine
$y_o$.

Having determined $T(y)$, we now determine the density field,
represented here by the separation $s(y)$. Inserting the solution of
$T(y)$ into Eq.~(\ref{eq:solts}), we obtain
\begin{equation}
\label{eq:tsy}
s(y) \simeq {t^2 d^2 \rho_o \over g} {B^2 v_w^2} {y_o \over (y_o - y)^2}
\exp (-2y / \ell),
\end{equation}
where $\rho_o$ is the density of the pile at rest. The separation $s$
also decays exponentially as one moves away from the bottom plate,
then increases for large values of $y$, similar to $T(y)$. Also, the
above expression for $s$ is not valid in the neighborhood of $y= y_o$.
Finally, we consider the expansion $y_{exp}$. The expansion can be
written as
\begin{eqnarray}
\label{eq:ts}
y_{exp} & \simeq & {1 \over d} \int _0^{y_o} s(y) dy \\ \nonumber
        & \simeq & {t^2 d \rho \over 2 g h_o} ~ {B^2 \ell v_w^2}.
\end{eqnarray}

\section{Discussion}
\label{sec:comp}

\subsection{Condition for the Gaseous State}
\label{sec:cgs}

In the previous section, we have obtained analytic results for the
vibrating box problem using Haff's theory. There are several
assumptions made in the construction of the theory as well as in the
derivation of the solution. In this section, we check the validity of
the solution, and in turn the assumptions, by comparing with the
results obtained by the MD simulations.

Before making the comparison, we check the condition under which the
system is regarded as gaseous. We consider that the system is gaseous
if the particles in the system stay afloat for the most part of time
colliding with each other. Therefore, for the gaseous system, the
vibration should provide enough input to make the particles float. In
other words, the collisional force caused by the bottom plate should
be large enough to overcome the pressure due to the gravity. Consider
a particle at the bottom of the box, where the pressure is the
greatest. The typical change of momentum during a collision of the
particle with the bottom plate is $m v_w$, where $m$ is the mass of
the particle, and $v_w$ is the typical velocity of the plate. Here, we
choose $v_w$ to be $2 \pi A f$, the maximum velocity of the plate.  We
then consider the frequency of such collisions. We argue that the
particle collides roughly once during one period, at least for the
small amplitudes of the vibration. It will be later shown that the
system becomes gaseous at a small value of the amplitude, so the above
estimation seems to be reasonable. The total force generated due to
the collisions is of order of $2 \pi m A f^2$. On the other hand, the
force on the particle due to gravity is of the order of $mgn$, where
$n$ is the number of layers in the pile. Then, the condition for the
gaseous system becomes that the ratio $F$ of the collisional force to
the gravitational force to be much larger than one. The ratio $F$ is
essentially the Froude number. The condition can be written as
\begin{equation}
F \sim {\Gamma \over 2 \pi n} \gg 1,
\end{equation}
where $\Gamma$ is $4 \pi^2 A f^2 / g$, the maximum acceleration of the
plate scaled by that of the gravity. Thus, the acceleration of the
bottom plate, not the velocity, is relevant in determining the
condition for the system to be gaseous.

We check the condition also by MD simulation. At a given time, we
measure the total number of pairs of particles in contact with each
other. We average the number over $200$ periods of the vibration.  The
system is gaseous, if the averaged number of pairs $N_p$ is much
smaller than the total number of particles $N$. Here, we use the
criterion that $N_p < 0.1 N$ for the system to be gaseous.  We choose
the width $W = 1$ and $N = 50$.  For several frequencies, we increase
the amplitude $A$ until $N_p / N$ becomes less than $0.1$. We obtain,
for each $f$, an interval of $A$ containing the value at which $N_p /
N = 0.1$. The intervals are $[0.06,0.08], [0.01,0.02]$ and
$[0.003,0.004]$ for $f = 20, 50, 100$, respectively. The corresponding
intervals of $\Gamma$ are $[0.97,1.29], [1.01,2.01]$ and
$[1.21,1.61]$. The acceleration at which the system becomes gaseous is
roughly constant, which is consistent with the above argument. We now
study the height dependence of the condition. We fix the frequency $f
= 100$ and the width $W = 1$, and vary $N$ to $100$ and $150$. By
using the same criterion, we find the intervals of $\Gamma$ to be
$[2.41,3.22]$ and $[4.03,5.64]$ for $N = 100$ and $150$, respectively.
The acceleration needed for the gaseous system is consistent with a
linear increase in $N$, which is again in agreement with the above
argument.  One thing to note is that $\Gamma$ to make the system
gaseous is fairly close $1$ for $N = 50$.  For the values of $\Gamma$
used in the simulations of the preceding section, the system is always
gaseous.  Also, $N_p$ for large amplitudes $A$ ($\sim 0.1$) is
essentially zero, which will be further discussed later.

\subsection{Scaling of Density and Temperature}
\label{sec:ef}

We compare the expansion $y_{exp}$ measured by the MD simulations with
that predicted by the theory. Measured values of $y_{exp}$, as shown
in Fig.~1, exhibit good scaling for the variable $x = A f$ with no
systematic deviation for the entire range of $x$. The scaling is
exactly what is predicted by the theory in Eq.~(\ref{eq:ts}). The
theory also predicts $y_{exp}$ is inversely proportional to the rest
height of the pile $H$ for fixed $A$ and $f$, which is certainly
consistent with the data presented in Fig.~5. On the other hand,
$y_{exp}$ should behave as $x^2$ according to the theory, but the data
(Fig.~1) is consistent with the behavior only for small values of $x$
(up to $\sim 2$). For larger values of $x$, $y_{exp}$ definitely
increases slower than $x^2$. The situations for the density $\rho (y)$
and granular temperature $T(y)$ fields are very similar. The value of
the fields at the bottom plate ($y=0$) shows good scaling for $x$
(Fig.~2), as again predicted in Eqs.~(\ref{eq:tsol}) and
(\ref{eq:tsy}).  The value of $T(0)$ from the simulations increases
for small $x$, reaches a maximum around $4$, than decays, very
different from the $x^2$ behavior given by Eq.~(\ref{eq:tsol}). The
behavior is again consistent with the prediction only for small values
of $x$. For the density field, we recall the relation between the
separation $s$ and the density $\rho$, which is
\begin{equation}
{s(0) \over d} \simeq ({\rho_o \over \rho (0)})^{1/3} - 1,
\end{equation}
where $\rho_o$ is the density of the rest pile. In Fig.~6, we plot
$\rho (0)^{-1/3}$ against $x^2$. The resulting curve, as predicted in
Eq.~(\ref{eq:tsy}), should be straight. The curve is indeed straight,
but only for small values of $x$. For large $x$, the curve increases
slower than linearly.

All three quantities we have studied behave in a similar manner.  They
all scale in the variable $x$, they are all consistent with the theory
for small values of $x$, and they all deviates from the predicted
behavior as $x$ is increased. Furthermore, the value of $x$ where they
start to deviate seems to be roughly the same ($x = 2 \sim 3$). We
recall that one of the assumptions in deriving the solution in the
previous section is quasi-incompressibility, which implies that the
separation $s$ is much smaller than $d$, the average diameter of the
particles. However, as shown in Fig.~2(a), the density of the pile
becomes the half of a packed density for $x \sim 3$, which is in
violation of the assumption. The system can not be treated as
quasi-incompressible for values of $x$ larger than about $3$. Since
the point where the quasi-incompressbility breaks down seems to
coincides to the point where the deviation from the theory starts, we
strongly suspect that the deviations are resulted from the breakdown.
Furthermore, if the deviation is due to the breakdown, it is clear why
the relevant variable for the breakdown is $x$, not $\Gamma$ or other
variable. The breakdown occurs when $s$ is comparable to $d$, and $s$
scales in the the variable $x$. Thus, the breakdown occurs at a
specific value of $x$.

We now consider the behavior of the entire density and temperature
fields. The density field $\rho (y)$, as shown in Fig.~3, roughly
scales in $x$ and has one maximum, both of which are consistent with
Eq.~(\ref{eq:tsy}). The temperature field $T(y)$ also scales in $x$,
but displays qualitatively different shapes (Fig.~4) depending on $A$.
For small $A$, the field has a local maximum around $y = 2$, while the
maximum disappears for larger $A$. On the other hand, the maximum is
always present in the prediction of the theory Eq.~(\ref{eq:tsol}). We
study the condition for the change of the shape. For given $f$, we
increase $A$ until the local maximum disappears. We obtain intervals
of $A$ containing the value at which the maximum disappears. They are
$[0.10,0.15], [0.04,0.06]$ and $[0.019,0.020]$ for $f = 20, 50, 100$,
respectively. The corresponding intervals of $x$ for the different
frequencies are roughly the same. We thus face a familiar
situation---the behavior of $T(y)$ is consistent with the theory for
small $x$, but deviates for large $x$.  Furthermore, the value of $x$
at which $T(y)$ deviates from the theory is about $2$, which is again
consistent with the previous argument.

\subsection{Beyond Incompressibility}
\label{sec:tcomp}

Since we argue quasi-incompressibility is valid only for modest range
of $x$, one might wonder why we do not relax the condition.  The
simple answer is that the system of equations becomes too complicated.
We now consider the changes needed in order to relax the
quasi-incompressibility. The three equations
Eqs.~(\ref{eq:mass})-(\ref{eq:energy}) are written in general form,
and do not need any modification. The relations of $p, \eta, K, I$ to
the fields Eqs.~(\ref{eq:intp})-(\ref{eq:I}) as well as the boundary
conditions Eq.~(\ref{eq:bc}) have to be modified. It is not these
modifications themselves, but the complexity of the resulting
equations, that makes the analytic solution too difficult to obtain. A
word about the regime of validity. The range of $x$ where the
quasi-incompressible theory is valid depends on $H$ the rest height of
the pile. For $H = 2$, which corresponds to the system of about $10$
layers, the theory is valid until $x$ reaches about $3$. As we
increase $H$, the range of validity also increases. Since a typical
experiment of granular system involves several tens or more layers,
the quasi-incompressible theory can describe quite large range of $x$,
and we hope many interesting phenomena occur within the regime of
validity.

One also might wonder why the scaling predicted by the theory is valid
even in the regime where the quasi-incompressible theory is no longer
valid. The only place where the external parameters $A$ and $f$ come
to the system is the boundary condition. The boundary condition, as
previously discussed, is a consequence of energy conservation. Even in
the general situation of a compressible system, the boundary condition
is obtained by equating two energy fluxes. One is the energy
transferred by collisions between the particles and the wall, which is
a function of the velocity of the wall. Now, the velocity is
proportional to $A f$, but the other flux term does not depend on $A$
or $f$, but on the fields of the system. Thus, it is not surprising to
see the scaling for the variable $A f$ holds even for the compressible
systems.

We finish the section by discussing two behaviors previously
mentioned. As noted in Sec.~\ref{sec:mexp}, the expansion $y_{exp}$
seems to increase faster near the largest value of $x$. This may be
explained as follows. As $A$ is increased for fixed $f$, the system
becomes more expanded, and the number of interparticle collisions
decreases. For sufficiently large $A$, it is be possible that a
particle in the system hardly interact with other particles, and it
mainly interacts with the bottom plate \cite{tmls89b}. This is
supported by the measurement of the average number of pairs $N_p$ in
contact, discussed in Sec.~\ref{sec:cgs}.  We find that $N_p$ is
essentially zero near the largest values of $A$. If we completely
ignore the interparticle interaction, it is straightforward to show
that $y_{exp} \sim x^2$.  Thus, $y_{exp}$ has to increase faster for
large $x$, which is consistent with the data (Fig.~1). However, it is
not clear, without further detailed study, that above mechanism is
actually operating.  Also, note that the mechanism predict scaling for
variable $x$. The behavior of $y_{exp}$ was shown to depend on
properties of the sidewall \cite{lhb94,lt}. It was shown that
$y_{exp}$ behaves as $x^{3/2}$ for entire range of $x$ with inelastic
sidewalls, while it deviate from the behavior with elastic ones (no
energy loss).

The other problem I want to discuss is the deviation from the scaling
for very small values of $f$. In Secs.~\ref{sec:mexp} and
\ref{sec:mfield}, we find nagging deviations
from the scaling of the expansion, the density and the temperature
field for $f = 20$ data. One possible way to explain them is to
consider the way the boundary condition is imposed. In constructing
the boundary condition, we assume that particles collide with the
bottom plate at random phase of the vibration. We thus assume the
``thermal motion'' is much larger than the convective motion. For low
enough frequencies, however, the particles have enough time to
dissipate their thermal fluctuations, so the assumption may not be
valid. In such cases, the convective motion of the particle dominates,
and one can no longer think of the system as particles with random
motion.

\section{Conclusion}
\label{sec:con}

We study the behavior of granular particles contained in a vibrating
box.  Using MD simulation, we show that several quantities which
describe the system---the density and the temperature fields as well
as the expansion---obey scaling in the variable $x = Af$. The behavior
of these quantities is qualitatively different for low and high values
of $x$. We also study the system using continuum equations developed
by Haff. We develop a boundary condition for moving boundaries, and
apply at the bottom plate of the box. We solve for the density and the
temperature fields of the time averaged steady state of the system.
Here, we are limited to the quasi-incompressible case, where the
average separation between the particles is much smaller than the
average diameter of the particles. The fields obtained from Haff's
equations show the same scaling as the simulational counterparts. The
behaviors of the fields from the theory are consistent with the
simulational data for small $x$, but they deviate significantly for
large $x$. We argue that the discrepancy is due to the fact that the
quasi-incompressibility condition we impose is not valid for large $x$.

In this paper, we have showed that Haff's theory, even with many
assumptions, seems to describe the system reasonably well in the
quasi-incompressible case (small $x$). It goes without saying that the
current study is only a very small step towards the understanding of
the system, and the following are what we consider important things
yet to be understood. The quality of the simulational data needs to be
improved in order to convincingly demonstrate the scaling. For
example, it is not certain from Figs.~3 and 4 that $\rho (y)$ and
$T(y)$ obey strict scaling or an approximate one. Also, the
improvement is necessary for quantitative comparison of the fields
from the theory with from simulations.  Next is the validity of Haff's
theory in the compressible regime (large $x$). As discussed in
Sec.~\ref{sec:tcomp}, due to the complexity of the problem in the
regime, it may be too difficult to obtain a closed form analytic
solution of the problem. However, with perturbative or approximate
methods, one might be able to get some information. Also, one can
numerically solve Haff's equations, and compare the results with MD
simulations. Another issue to be understood are time dependent
properties of the system. In this paper, we have studied only steady
state properties, which averages out the variations of the fields
during one period of the vibration. Time dependent properties can be
studied by perturbation from the steady state \cite{td}.  And of
course, Haff's theory has to be applied to other systems to find the
power and limitation of the theory.

The author thanks Joel Koplik and Stefan Luding for many useful
discussions and comments on the manuscript. This work is supported in
part by the Department of Energy under grant DE-FG02-93-ER14327.

\newpage
\section*{Figure Captions}

\begin{description}

\item [Fig.~1:] The expansion $y_{exp}$ for several
values of the amplitudes $A$ and the frequencies $f$ of the vibration.
We find the expansion exhibits good scaling behavior with variable $A
f$.

\item [Fig.~2:] (a) The density and (b) the square root of
granular temperature at the bottom plate for various values of $A$ and
$f$ of the vibration. Here, we find that the same scaling as the
expansion holds.

\item [Fig.~3:] The density field for various values of $A$ and $f$.
Here, value of $A f$ is fixed to be (a) $3$ and (b) $10$. We find that
the same scaling as the expansion $y_{exp}$ seems to hold.  However,
the quality of scaling is poorer.

\item [Fig.~4:] The square root of the granular temperature field
for various values of $A$ and $f$. The value of $A f$ is fixed to be
(a) $3$ and (b) $10$. We again find poorer scaling similar to the
density field (Fig.~3).

\item [Fig.~5:] The scaled expansion vs $A f$ for $N = 50, 100, 150$ and
for several values of the amplitude. All the curves seems to collapse
to a single curve.

\item [Fig.~6:] Plot of $\rho (0)^{-1/3}$ against $x^2$ calculated from
Fig.~2(a). The curve is straight for small values of $x$, and deviates
from the linear behavior for large $x$.

\end{description}


\begin{thebibliography}{99}
\bibitem{timestep} The expansion as well as other quantities are also
measured with reduced timestep in order to check timestep is
sufficiently small. For a given $f$, quantities for the largest $A$ is
measured with a timestep being reduced by factor of at least $2$, and
in some cases by $5$ or $10$. We find no essential difference for
smaller timesteps.
\bibitem{t} In Ref.~\cite{h83}, the temperature $T$ is replaced by
${\bar{v}}^2$. It is simply a matter of notation, and does not
change any of the results.
\bibitem{tmls89b} In Ref. \cite{tmls89}, the state of no interparticle
interaction is observed in the experiments, and is named as
``Newtonian state.''
\end{thebibliography}
\end{document}